\documentclass[lettersize,journal]{IEEEtran}
\usepackage{amsmath,amsfonts}
\usepackage{algorithmic}
\usepackage{algorithm}
\usepackage{array}
\usepackage[caption=false,font=normalsize,labelfont=sf,textfont=sf]{subfig}
\usepackage{textcomp}
\usepackage{stfloats}
\usepackage{url}
\usepackage{xcolor}
\usepackage{verbatim}
\usepackage{graphicx}
\usepackage{cite}
\usepackage{xcolor}
\usepackage{tabularx,booktabs,textcomp}
\usepackage{amsthm}

\usepackage{amsmath,amssymb,amsthm}

\begin{document}
\title{DynamoPMU: A Physics Informed Anomaly Detection and Prediction Methodology using non-linear dynamics from $\mu$PMU Measurement Data}

\author{{Divyanshi Dwivedi,~\IEEEmembership{Student Member,~IEEE}, Pradeep Kumar Yemula,~\IEEEmembership{Member,~IEEE}, Mayukha Pal,~\IEEEmembership{Senior Member,~IEEE}}
 

\thanks{(Corresponding author: Mayukha Pal)}

\thanks{Mrs. Divyanshi Dwivedi is a Data Science Research Intern at ABB Ability Innovation Center, Hyderabad 500084, India and also a Research Scholar at Department of Electrical Engineering, Indian Institute of Technology, Hyderabad 502205, IN, (e-mail: divyanshi.dwivedi@in.abb.com).}

\thanks{Dr. Mayukha Pal is a Global R\&D Leader – Data Science at ABB Ability
Innovation Center, Hyderabad-500084, IN, (e-mail: mayukha.pal@in.abb.com).}

\thanks{Dr. Pradeep Kumar Yemula is an Assoc. Professor with the Department of Electrical Engineering, Indian Institute of Technology, Hyderabad 502205, IN, (e-mail: ypradeep@ee.iith.ac.in).}
}

\maketitle

\begin{abstract}

The expansion in technology and attainability of a large number of sensors has led to a huge amount of real-time streaming data. The real-time data in the electrical distribution system is collected through distribution-level phasor measurement units referred to as $\mu$PMU which report high-resolution phasor measurements comprising various event signatures which provide situational awareness and enable a level of visibility into the distribution system. These events are infrequent, unschedule, and uncertain; it is a challenge to scrutinize, detect and predict the occurrence of such events. For electrical distribution systems, it is challenging to explicitly identify evolution functions that describe the complex, non-linear, and non-stationary signature patterns of events. In this paper, we seek to address this problem by developing a physics dynamics-based approach to detect anomalies in the $\mu$PMU streaming data and simultaneously predict the events using governing equations. We propose a data-driven approach based on the Hankel alternative view of the Koopman (HAVOK) operator, called DynamoPMU, to analyze the underlying dynamics of the distribution system by representing them in a linear intrinsic space. The key technical idea is that the proposed method separates out the linear dynamical behaviour pattern and intermittent forcing(anomalous events) in sequential data which turns out to be very useful for anomaly detection and simultaneous data prediction. We demonstrate the efficacy of our proposed framework through analysis of real $\mu$PMU data taken from the LBNL distribution grid. DynamoPMU is suitable for real-time event detection as well as prediction in an unsupervised way and adapts to varying statistics.

\end{abstract}


\section{Introduction}
\label{section:Introduction}
\subsection{Background and Motivation}

The modern electrical distribution grid is equipped with many utility assets and equipment which include capacitor banks, tap-changing transformers, protection devices, dynamic loads, electric vehicles, and distributed energy resources \cite{dd}. The normal and abnormal switching operations of these assets can create various events in the system, and the behaviour of these events has different signature patterns \cite{graphpmu}. It is an important aspect of capturing and detecting the events; allows awareness and enhances the visibility of the distribution grid \cite{situational}, helps in diagnosing the health of the asset \cite{asset1}, analyzes the distribution-level oscillation occurring because of distributed energy resources \cite{locationpmu}, microgrid synchronization \cite{syn}, and fault detection and analysis \cite{faultpmu}.
\begin{figure}
  \centering
  \includegraphics[width=3.3in]{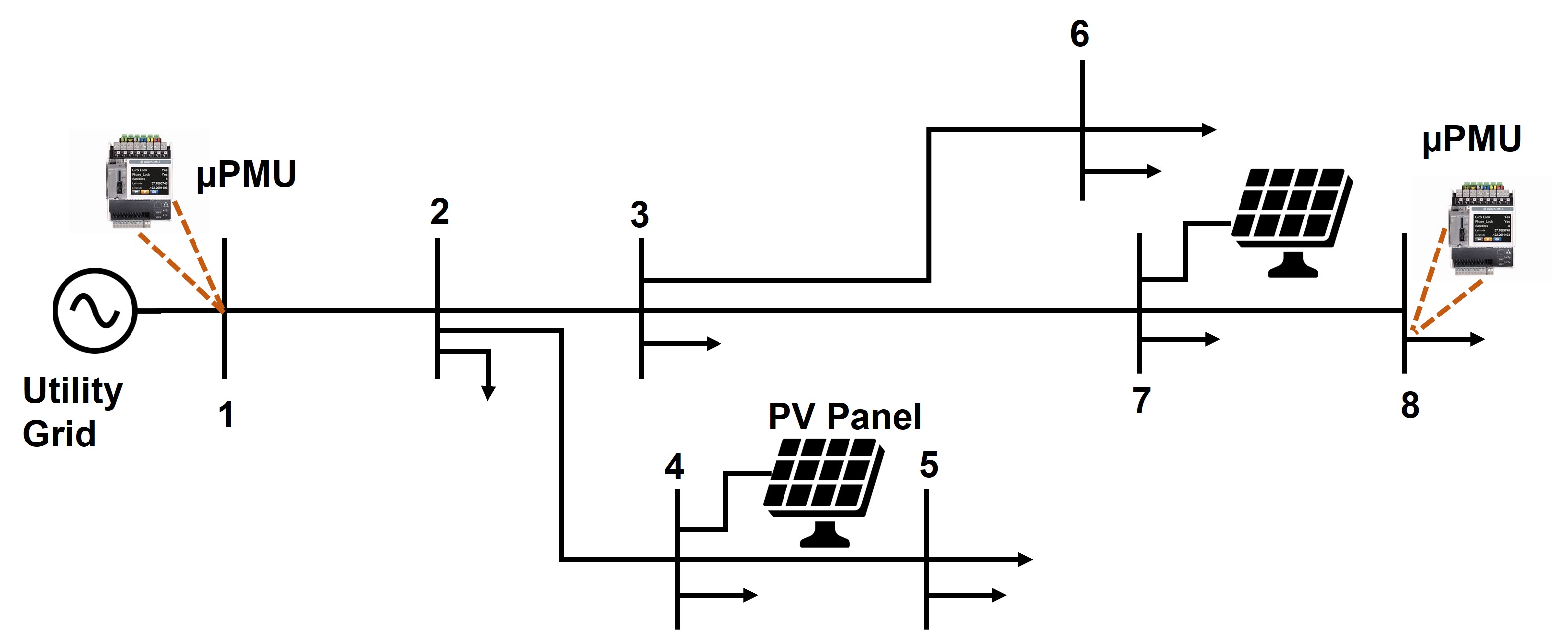}\\
  \caption{$\mu$PMU installed in 8-bus distribution feeder system.}
  \label{fig:distribution}
\end{figure}

\begin{figure}
  \centering
  \includegraphics[width=3.2in]{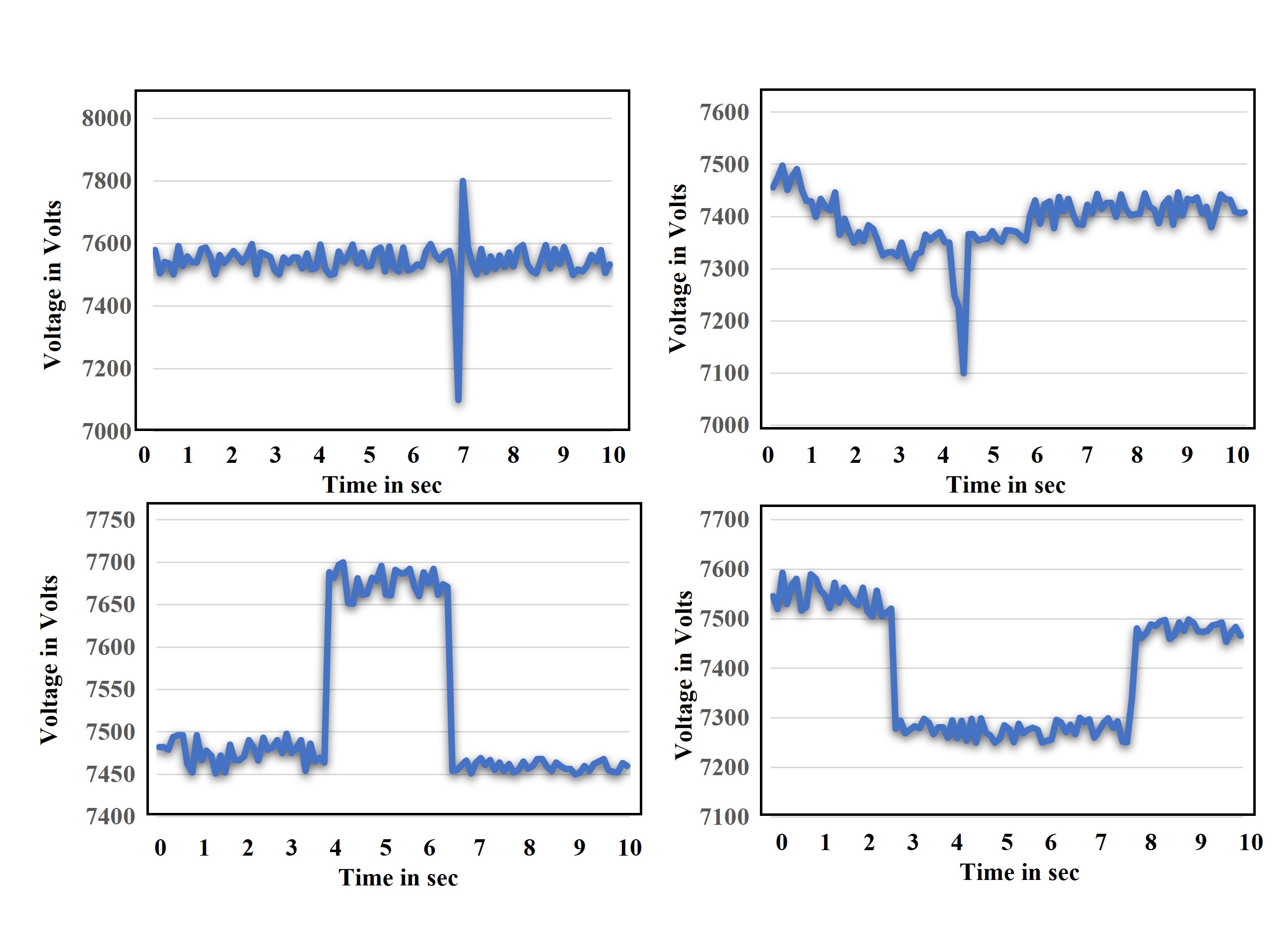}\\
  \caption{Example of anomalies detected in voltage signal.}
  \label{fig:voltage_anomaly}
\end{figure}

These analyses and event detection have been made possible with the integration of distribution-level phasor measurement units in smart grids named $\mu$PMU \cite{ggl} as shown in Fig. \ref{fig:distribution}. $\mu$PMU provides all three-phase measurements of voltage and current magnitude and phase angle at 120 readings per second. Using the enormous measurements of $\mu$PMU, data-driven techniques were found suitable as they automate the process of detection as well as prediction of events. The data-driven techniques help in identifying infrequent, uncertain and unscheduled occurring events. The example of anomalous events in voltage signal is shown in Fig. \ref{fig:voltage_anomaly}. In this work, we solve this dynamic problem of detecting anomalies in the $\mu$PMU data and simultaneously predicted the events occurring in the long term by using physics informed unsupervised data-driven method.

\subsection{Summary of Technical Contributions}
We use the online available real-world dataset of $\mu$PMU measurement which enables advanced diagnostic and monitoring strategies in distribution systems. The dataset involves various unknown, infrequent and unscheduled natures of events. We propose a physics-inspired dynamic approach which analyzes the measurement data without the requirement of prior knowledge about the event signatures. The key contributions of this paper are listed as follows:

\begin{itemize}
    \item A novel physics-inspired analytics method which represents the nonlinear dynamical electrical distribution system by a linear model based on the Hankel matrix alternative view of Koopman (HAVOK) theory. For detecting and analyzing the change in dynamical behaviour (i.e., anomalous events) in various linear inherent coordinates we identify the intermittent forcing in the dynamic system. The proposed method has the capability to represent complex dynamics and effectively simplify the dynamical pattern through the developed framework. The $\mu$PMU could easily be analyzed based on this dynamic system theory without making any prior knowledge about the system which shows the robustness to handle the data-inherent uncertainty. \\
    \item It involves time-delay embedding theory and dynamic mode decomposition (DMD) combined with the Koopman operator; a linear infinite-dimensional operator. The finite approximation of the Koopman operator is the key phenomenon which evolves the inherent functions to acquire a similar dynamical behaviour as the original
    nonlinear dynamic response through the Koopman transformation. From that, we identify the different dynamical signatures or events in the data.\\
    \item Accordingly, any dynamical signatures or patterns that deviate from the linear dynamical response are considered as the event in the $\mu$PMU data. The proposed method analyses the dynamical behaviour in the $\mu$PMU data thus we named it DynamoPMU. To check the effectiveness of the proposed event detection method compared to state-of-the-art methods in the literature. We computed various metrics scores which show that it outperforms the existing methods. This method only requires the $\mu$PMU measurement data without any information about the network model or prior labelling of the events.\\
    \item Using the governing equations obtained while performing the anomaly detection, we can predict the $\mu$PMU measurements and anomalies simultaneously for the long term. This is achieved by using a sparse identification of nonlinear dynamics (SINDy), which prominently predict the measurements by considering the dynamical governing equations of the system. To train the model we use 10 days of $\mu$PMU data and predicted the measurements for the next two days. The obtained results give better metrics scores for the prediction which shows the reliability of the proposed method.

\end{itemize}

\subsection{Literature Review}

Event detection has been explored significantly in the electrical distribution grid. In literature, data-driven methods have been proposed based on statistical methods \cite{stat2}, machine learning-based supervised, semi-supervised and unsupervised learning algorithms. Statistical methods involve the concept of absolute deviation around the median with consideration of dynamic window sizes \cite{situational}. Using null and alternative hypothesis test tools of statistics frequency deviation, including load variation and faults is detected and further detected high impedance fault (HIF) which is the hardest to detect \cite{stat1}. In terms of metrics computed for event detection, the statistical method lacks performance in comparison to the proposed DynamoPMU method. 

The implementation of machine learning models got extensive attention because of the great amount of measurement data available. Using supervised and semi-supervised learning which requires either full labelling or partial labelling of the events has been implemented in \cite{supervised1}, \cite{supervised2}, \cite{supervised3} and many others. However, it is difficult to achieve prior labels for the events, thus the implementation of these methods is not suitable for real-world applications. Further, researchers also acknowledge the implication of deep learning unsupervised learning algorithms to overcome the gap of requirements of prior labelling. Available work focuses on detecting some particular events, such as frequency-related events \cite{frequency}-\cite{frequency1}, fault events \cite{fault1}, capacitor bank switching \cite{capacitor} or voltage-related events \cite{event2}. However, we propose an event detection method which covers the detection of a wide range of events. There are research papers available which can also detect a wide range of events using unsupervised deep learning algorithms which includes the implementation of Generative Adversarial Networks (GAN) \cite{event1}, \cite{event3}. We have compared the performance of the method in terms of metrics and our proposed method shows better results for detecting the events. Another unsupervised method based on ensemble learning is proposed to develop a model for fast, scalable bad data/event detection for PMU \cite{event4}. These methods are suitable for offline event detection while the proposed method has the capability to be implemented for real-time streaming data. The proposed method has the capability to detect the events ahead of their actual occurrence and this could be utilized as an alarm signal to the operator for situational awareness. Further, this approach using the past data simultaneously can predict the measurements as well as anomalies for the long term. In \cite{htm}, they worked on real-time anomaly detection as well as simultaneous prediction. This method based on hierarchical temporal memory has the capability to learn online in one pass and adapt to changing statistics, but it cannot detect prior to the event occurrence and it has not detected many signature events. Thus, DynamoPMU analyzes the changes in the dynamic behaviour of the measurement data which allows it to capture various events significantly prior to their occurrence.

The main approach used in the paper for anomaly detection is inspired by the HAVOK analysis, it has been utilized broadly in various areas such as biomedical signal processing for detecting anomalies in multi-variate EEG system \cite{eeg}, pathophysiological processes of obstructive sleep apnea \cite{patho}, real-time control of robotic systems \cite{robotics} and turbulence flow study \cite{turbulence}, \cite{turbulence1}. The method works efficiently in these fields thus considering its performance in other domains we utilized its implementation for detecting anomalies in the electrical distribution system.

For accurate prediction, the implementation of SINDy has been found very effective in predicting the transmission dynamics of COVID-19 \cite{covid}, discovering mechanistic equations for measles, chickenpox, and rubella \cite{disease}, accurate prediction of thermal comfort in the vehicle cabin \cite{vehicle}, prediction of blood glucose \cite{blood} and many others. It has also been used in the power grid to predict online voltage evolution to replace the static voltage-sensitivity analysis \cite{voltage}. SINDY models have shown a superior performance when dealing with highly nonlinear, high-dimensional, multi-scale dynamical systems with limited training data and low computational burden.

The paper is organized as follows: a description of the methodology is detailed in Section \ref{section:Material} with the process of how we detected the anomalies and predicted the measurements. Section \ref{section:Result} provides the analysis and discussion for the obtained results. Finally, Section \ref{section:Conclusion} concludes the paper.

\section{Materials and Methods}
\label{section:Material}

The integration of renewable energy sources, electric vehicles, demand response, and peer-to-peer energy trading changes the network's load profiles and configuration spontaneously. This complex interaction causes great uncertainties, and even bidirectional power flow in the distribution network, making distribution networks' behavioural response more non-linear and dynamic. The proposed framework is found to be suitable to analyze the dynamic response and understand the non-linear behaviour of the electrical distribution system. Let us consider a dynamic electrical distribution system in the following form:

\begin{equation}
\label{eq:dyn}
\frac{d}{dt} x(t) = f(x(t))
\end{equation}

where $x(t) \in \mathbb{R}^n$ is the state of system at time $t$, $f$ is a function which describe the dynamics of the state $x$. Equation 1 represents a continuous-time dynamical system, while taking $\mu$PMU data in samples it is represented as:

\begin{equation}
x_{k+1}=F(x_k)= x_k + \int_{k \Delta t}^{(k+1) \Delta t}  f(x(\delta))d\delta
\end{equation}

where, $x_k= x(k \Delta t)$ represents the system's trajectory from equation \ref{eq:dyn} and $F$ is a discrete-time propagator.

\subsection{Koopman Operator Theory}

Traditionally, the geometric outlook of a dynamic system transverse trajectory which are dependent on fixed points, periodic orbits and attractors. On the other hand, a Koopman operator introduced in 1931, provides a better and alternative view which analyzes the measurements of state $y=m(x)$. It is the more practical view for analyzing a nonlinear complex dynamical system through observation space, as measurement data is growing abundantly. Koopman operator $(\mathbb{K})$ is a linear representation of a nonlinear system by converting the n-dimensional state vector to an infinite dimensional state vector. Thus, it is defined as an infinite-dimensional linear operator that works on measurement functions $m:G\rightarrow \mathbb{R}\ $ of state $x$ and given as:

\begin{equation}
    \mathbb{K}_m= m \circ F
\end{equation}

where, $\circ$ is a composition operator, which can be expanded as:

\begin{equation}
    \mathbb{K}_m(x_k)= m(F(x_k)) = m(x_k+1)
\end{equation}

\begin{figure}
  \centering
  \includegraphics[width=3.2in]{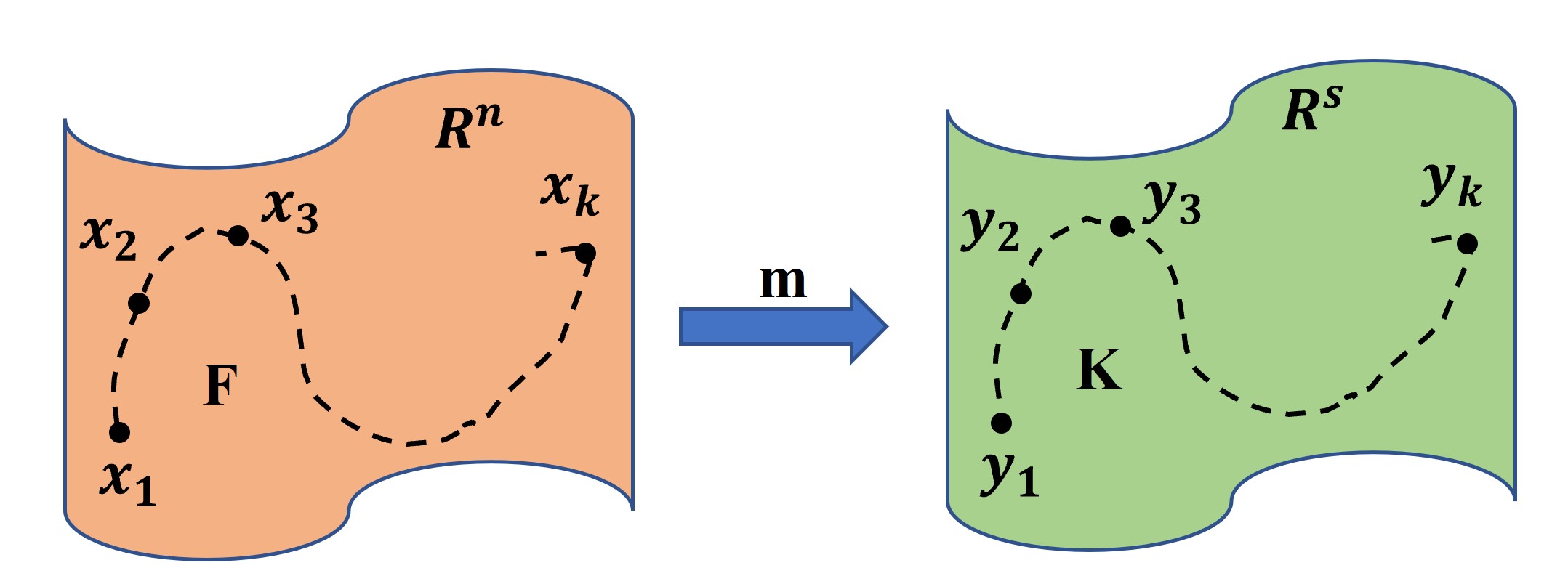}\\
  \caption{Koopman operator linearizing a nonlinear system using an observable measurement function $m$.}
  \label{fig:meas}
\end{figure}

The measurements in the given space are:

\begin{equation}
    m= \alpha_1 m_1 + \alpha_2 m_2 + \dots + \alpha_s m_s
\end{equation}

after applying the Koopman operator, $m$ remains in the same state space and is given as:

\begin{equation}
    \mathbb{K}_m= \beta_1 m_1 + \beta_2 m_2 + \dots + \beta_s m_s
\end{equation}

where, $\alpha_s$ , $\beta_s \in \mathbb{R}$, $\forall s= 1,\dots,s$ are the coefficients for linear subspace for the observations $m_s$ before and after application of Koopman operator for $s-$dimensional measurement subspace. The linear system that supports the Koopman observable is bounded to the subspace as:

\begin{equation}
    y_{k+1}= \mathbb{K} y_k
\end{equation}
\begin{equation}
    y_k= [m_1(x_k), m_2(x_k), \dots , m_s(x_k)]^T
\end{equation}

where $y_k$ is a measurement vector in the Koopman invariant subspace as shown in Fig.\ref{fig:meas}. 

Thus, the Koopman operator expresses a nonlinear system into a linear system. However, finite-dimensional approximation of the Koopman operator is challenging \cite{havok}. Using dynamic mode decomposition (DMD), a best-fitted linear matrix which represents spatial measurements is implemented, but it possesses a large error in many nonlinear systems \cite{dmd}. Another extended DMD was proposed but that is not reliable \cite{dmdext}. To overcome the drawback, intrinsic data-driven measurement coordinates are derived from the time history of the system measurement, namely eigen time-delay coordinate is proposed \cite{havok}. In this work, we have utilized the eigen time-delay coordinates by satisfying Taken's embedding theorem conditions.

\begin{figure*}
  \centering
  \includegraphics[width=7.0in]{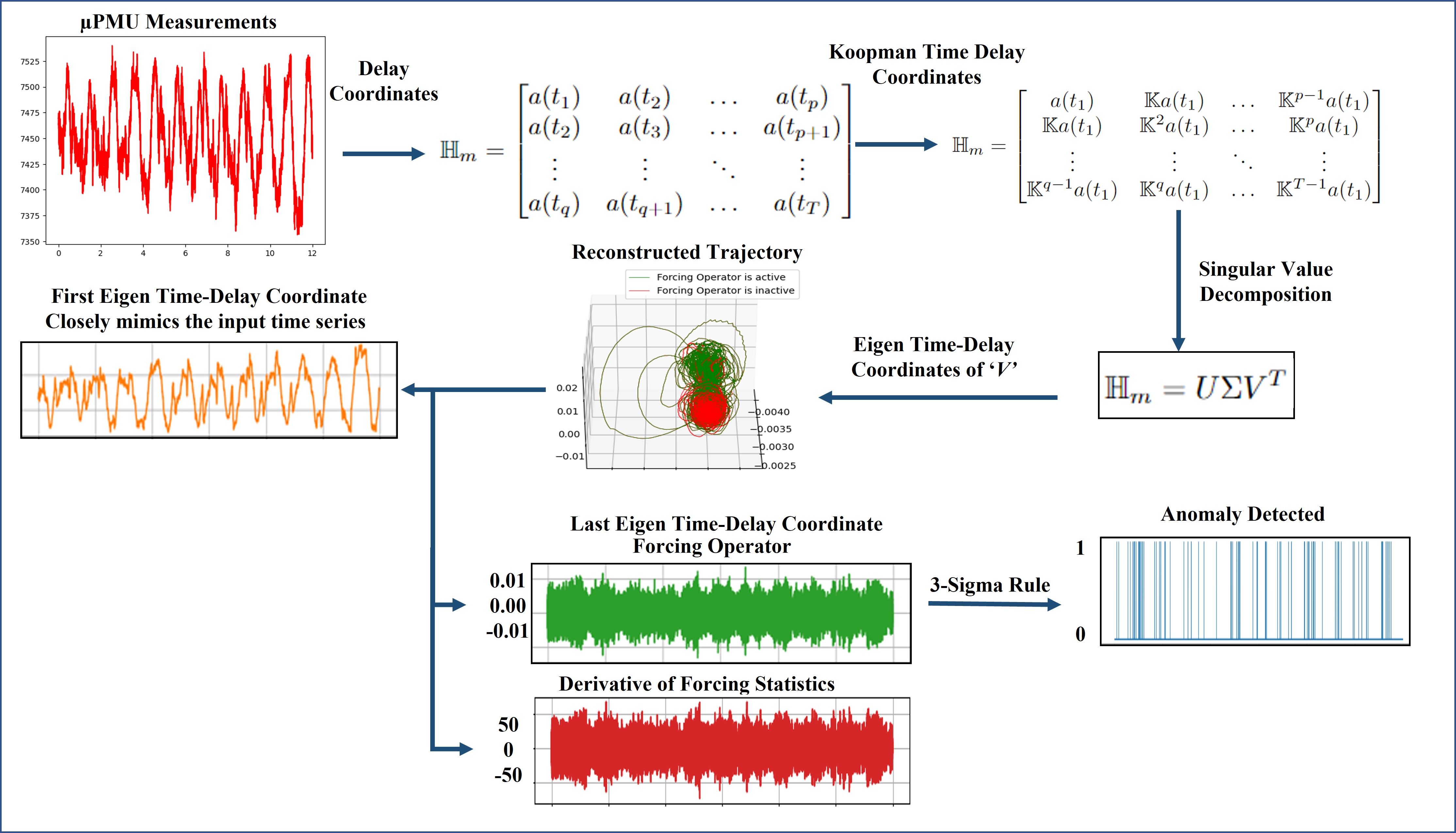}\\
  \caption{An illustration of the proposed Hankel alternative view of Koopman operator-based analytics framework for anomaly detection and prediction in $\mu$PMU measurements. The measurements with delay coordinates are used to create a Hankel matrix which is further modified in terms of the Koopman operator and then Singular value decomposition is applied. From the achieved components, eigen time-delay coordinates of $``V"$ are analyzed by separating the first eigen time-delay coordinates $V_{r-1}$ that mimic the pattern similar to the input time series by providing the best-fit linear dynamical model and the last eigen time-delay coordinate $V_{r-1}$ is considered as stochastic input that intermittently forces the first ${r-1}$ variables. Here, we decomposed the measurement data into linear dynamics with intermittent forcing terms.}
  \label{fig:method}
\end{figure*}

\subsection{Takens’ embedding theorem for state space reconstruction}

Takens’ embedding theorem suggests that we can enhance the measurements $a(t)$ with time shifts $a(t-\tau)$ known as delay coordinates such that the attractor of a dynamical system is an isomorphism of smooth manifolds to the original attractor under specific conditions. In accordance with the theorem, the trajectory is formed by points $\hat{x_i}$ using the delay map as follows:
\begin{equation}
    \hat{x_i}= [p(x_i), p(x_{i+\tau}), \dots , p(x_{i+(m-1)\tau})]
\end{equation}
\begin{equation}
    \hat{x_i}= (a_i,a_{i+\tau}, \dots ,a_{i+(m-1)\tau})
\end{equation}
\begin{equation}
    i= 1, \dots , N-e+1
\end{equation}

where $\tau$ is the time lag, $e$ is the embedding dimension and $N$ length of time-series.

\subsection{Hankel alternative view of Koopman (HAVOK) analysis}

The measurement time-series $a(t)$ from $\mu$PMU is taken and Hankel matrix $\mathbb{H}_m$ is formed under the assumption that the conditions of the Taken’s embedding theorem are satisfied \cite{koopman}. Then we compute eigentime delay coordinates by performing singular vector decomposition (SVD) of the Hankel matrix $\mathbb{H}_m$.

Hankel Matrix, \begin{equation}
\mathbb{H}_m= \begin{bmatrix}
a(t_1) & a(t_2) & \dots & a(t_p) \\ 
a(t_2) & a(t_3) & \dots & a(t_{p+1}) \\ 
\vdots & \vdots & \ddots & \vdots \\ 
a(t_q) & a(t_{q+1}) & \dots & a(t_T) \\ 
\end{bmatrix} 
\end{equation}

where, time points $t_{i+1}= t_i+\tau (i=1, \dots, T-1)$, $q$ represents the number of points in the trajectory and $p$ represents window length which means the longest periodicity attained by the Hankel matrix. After implementing SVD we get,
\begin{equation}
    \mathbb{H}_m= U \Sigma V^T
\end{equation}

The columns from SVD obtained are $U$ and $V$ which are exhibited hierarchically in decreasing order. Generally, $\mathbb{H}_m$ acquires a low-rank approximation for the first $r-th$ columns of $U$ and $V$ which raise a measurement subspace that is invariant to the Koopman variant for the states. Thus we can rewrite the Hankel matrix as:

\begin{equation}
\mathbb{H}_m= \begin{bmatrix}
a(t_1) & \mathbb{K}a(t_1) & \dots & \mathbb{K}^{p-1}a(t_1) \\ 
\mathbb{K}a(t_1) & \mathbb{K}^2a(t_1) & \dots & \mathbb{K}^pa(t_1) \\ 
\vdots & \vdots & \ddots & \vdots \\ 
\mathbb{K}^{q-1}a(t_1) & \mathbb{K}^qa(t_{1}) & \dots & \mathbb{K}^{T-1}a(t_1) \\ 
\end{bmatrix} 
\end{equation}

Hankel matrix's rows and columns are well estimated by the first $r-th$ columns and rows of $U$ and $V$ respectively, which are referred to as eigen time-series which provides a Koopman-invariant measurement system. The first $r$ columns of $V$ give a time series of the magnitude for each of the columns of $U\Sigma$ in the data. By plotting the first three columns of $V$, we obtain an embedded attractor as shown in Fig. \ref{fig:method}. The SINDy algorithm is applied to achieve a forced linear system from the delay coordinates and obtain a good linear fitting for the initial $r-1$ variables and a bad fit for the $r-th$ variable. Basically, eigen time-delay with the Koopman operator performs a linear regression where a linear model built on the first $r-1$ variables in $V$ and $v_r$ as an intermittent forcing term which is given as:

\begin{equation}
    \frac{d}{dt}v(t) = Av(t) +Bv_r(t)
\end{equation}

\begin{equation}
    \frac{d}{dt} \begin{bmatrix}
        v_1\\
        v_2\\
        \vdots\\
        v_{r-1}\\
        v_r
    \end{bmatrix}= \begin{bmatrix}
        A & B\\
        0 & 0\\
    \end{bmatrix} \begin{bmatrix}
        v_1\\
        v_2\\
        \vdots\\
        v_{r-1}\\
        v_r
        \end{bmatrix}
\end{equation}

where $v=[v_1, v_2, \dots, v_{r-1}]^T$ represents the first $r-1$ eigen time-delay coordinates. $v_r$ represented as a forcing in the linear dynamical system means nonlinear dynamics in the system. The statistics of $v_r(t)$ are non-Gaussian in nature, its long tail corresponds to rare events that drive a lobe switching in the dynamical system. Although statistics of $v_r(t)$ is not sufficient in identifying the high-frequency bursts, thus dynamics splitting concept is applied to chaotic systems, in which the Koopman operators have a continuous spectrum which is given as:
\begin{equation}
    V_r=\begin{bmatrix}
        v^{(1)} & v^{(2)} & \dots & v^{(q-1)}\\
        v^{(1)}_r & v^{(2)}_r & \dots & v^{(q-1)}_r\\
    \end{bmatrix}
\end{equation}
\begin{equation}
    V_r^{'} =\begin{bmatrix}
        v^{(2)} & v^{(3)} & \dots & v^{(q)}\\
        v^{(2)}_r & v^{(2)}_r & \dots & v^{(q)}_r\\
    \end{bmatrix}
\end{equation}

\begin{equation}
    \mathbb{A} \approx = V_r^{'} V_r^*
\end{equation}

where, $V_r^{'}$ is the 1-step time advanced eigen time-delay coordinates of $V_r$ and $V_r^*$ is the pseudo inverse of $V_r$ computed using SVD.
The matrices $V_r$ and $V_r^{'}$ are linkable by a best fitting operator $\mathbb{A}$ which minimize the Frobenius norm error $||V_r^{'}- \mathbb{A}V_r||$. Further, sparse identification of nonlinear dynamical systems (SINDy) \cite{SINDy} is implemented to obtain the first eigen decomposition of $\mathbb{A}$. 
Finally, the thresholding is applied to the values of forcing operators using the three-sigma rule \cite{3sigma}, which states that if values go beyond $\mu \pm 3\sigma$, those values are considered anomalies in the measurements.

\subsection{SINDY-based measurements and anomaly prediction}

From the available measurements of $\mu$PMU dynamic data set of the distribution system, the SINDY
identifies fully non-linear dynamical terms using a modified version of equation \ref{eq:dyn} given as:
\begin{equation}
    \Dot{X} = \Theta^{T} (X)
\end{equation}

where $X$ is a Hankel matrix given in equation 12. The term $\Theta^{T}$ is a library of the candidate dynamics which enhances the performance of the SINDY algorithm \cite{Sindy1}. In the electrical distribution system, the library of candidate dynamics is chosen which comprises a mixture of nonlinear trigonometric functions such as $\sin{(\delta - \theta)}$ or $\cos{(\delta - \theta)}$.  Using a sparse regression algorithm, sparse coefficients $\zeta_k$ are obtained from the initial values $\hat{\zeta_k}$ given as:
\begin{equation}
    \zeta_k= argmin \zeta_k ||\Dot{X}_k-\hat{\zeta_k} (X) \Theta^{T}||_2 + \alpha ||\hat{\zeta_k}||_1
\end{equation}

where, $\Dot{X}_k$ represents the $k-th$ row of $\Dot{X}$. The value of $\alpha$ is chosen such that the prediction model maintain complexity and accuracy.

\subsection{Source of $\mu$PMU datasets}

For anomaly detection and prediction in $\mu$PMU measurements using the proposed DynamoPMU, we use an open-source dataset provided by Lawrence Berkeley National Laboratory’s (LBNL) distribution grid \cite{pmudata}. This is the first dataset of $\mu$PMU installed in an electrical distribution grid which is available for research. We chose a dataset of $\mu$PMU installed in a $7.2$ kV grid, named a6bus1. The used $\mu$PMU is PQube3 which measures the three-phase voltage and current phasor and magnitudes at 512 samples per cycle,  60  cycles per second. The $\mu$PMUs measure phase angle and output synchrophasor data at 120 Hz frequency. The available data is in millisecond resolution for analysis we have re-sampled at second intervals for 12 days and 13 hours (data points 1,083,600). The events that occurred are of long duration, thus we can easily capture them in resampled data. The voltage and current magnitude are considered to detect the anomaly, however, the phase angles of voltage and current are often not used directly because of leading frequency fluctuations. Thus, using voltage and current angle measurements we computed the power factor and considered it as a feature \cite{event1}.
\subsection{Evaluation Metrics}

For validating the performance of the proposed DynamoPMU to detect and predict the anomalies in a dynamic electrical distribution system. Metrics for analyzing the performance of the proposed algorithm for detecting anomalies are given:

\begin{figure*}
  \centering
  \includegraphics[width=7.0in]{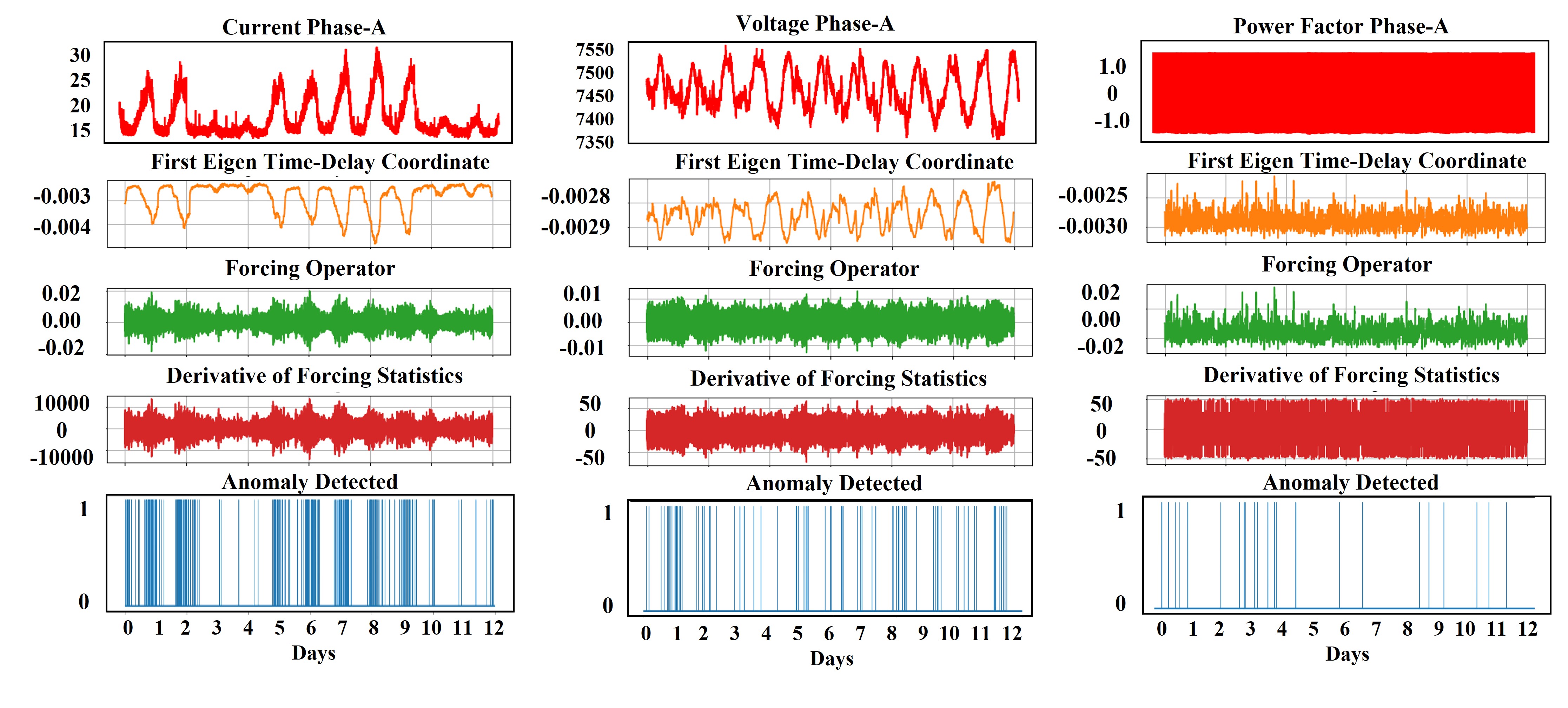}\\
  \caption{Current, Voltage and Power Factor measurements fed to the proposed DynamoPMU, obtained the forcing operator which suggests the intermittency in the measurements, further implemented the 3-sigma rule to detect the anomalies.}
  \label{fig:anomaly}
\end{figure*}

\begin{itemize}
    \item Precision- Evaluate the fraction of correctly classified events among the positively classified events.
    \begin{equation}
        Precision = \frac{TP}{(TP + FP)}
    \end{equation}
    \item Recall- Evaluate total relevant events classified correctly.
    \begin{equation}
        Recall = \frac{TP}{(TP + FN)}
    \end{equation}
    \item F1 Score- It is a harmonic mean of the precision and recall.
    \begin{equation}
        F1 Score = \frac{2 * Precision * Recall}{(Precision + Recall)}
    \end{equation}
    \item Accuracy- calculated in terms of positives and negatives as follows:
    \begin{equation}
        Accuracy = \frac{TP+TN}{(TP+TN+FP+FN)}
    \end{equation}
    \item Area under the ROC Curve (AUC)- It is an average of true positive rates over all possible values of the false positive rate.
     \item Matthews’s Correlation Coefficient (MCC)- It is the most reliable statistical metric which provides a good score only if the event detection is appropriate in all of the four measures of the confusion matrix. Mathematically, it is expressed as:
\begin{multline}
    MCC= \\ \frac{(TP \times TN-FP \times FN)}{\sqrt{(TP+FP)(TP+FN)(TN+FP)(TN+FN)}}
\end{multline}

where, TP = True Positives, TN = True Negatives, FP = False Positives, and FN = False Negatives.
\end{itemize}

Metrics for analyzing the performance of the proposed algorithm for predicting anomalies are given:

\begin{itemize}
    \item $R^2$ Score- It is known as the coefficient of determination, which measures how well a statistical model predicts an outcome. The best possible score is 1.0 and calculated as:
    \begin{equation}
        R^2(y,\hat{y})= 1- \frac{\sum_{i=1}^{n}(y_i-\hat{y_i})}{\sum_{i=1}^{n}(y_i-\Bar{y_i})}
    \end{equation}
    $\hat{y_i}$ is the predicted value, $y_i$ is the actual value of $i-th$ sample and $\Bar{y_i}$ is the mean of true values.
    \item RMSE-  Root-mean-square error is a standard deviation of prediction errors or residuals. It is computed as:
    \begin{equation}
    RMSE(y,\hat{y})= \frac{1}{N_s}\sum_{i=0}^{N_s-1}(y_i-\hat{y_i})^2
    \end{equation}
    where, $N_s$ is the number of samples.
    \item Explained variance score- It explains the dispersion of errors of a given dataset. The best possible score is 1.0 and computed as:
    \begin{equation}
        Variance (y,\hat{y})=1-  \frac{Var(y-\hat{y})}{Var(y)}
    \end{equation}
     \item MAE- The mean absolute error function computes a risk metric corresponding to the expected value of the absolute error loss or $l$-norm loss.
    \begin{equation}
        MAE(y,\hat{y})= \frac{1}{N_s}\sum_{i=0}^{N_s-1}|y_i-\hat{y_i}|
    \end{equation}
\end{itemize}

\begin{figure*}
  \centering
  \includegraphics[width=5.8in]{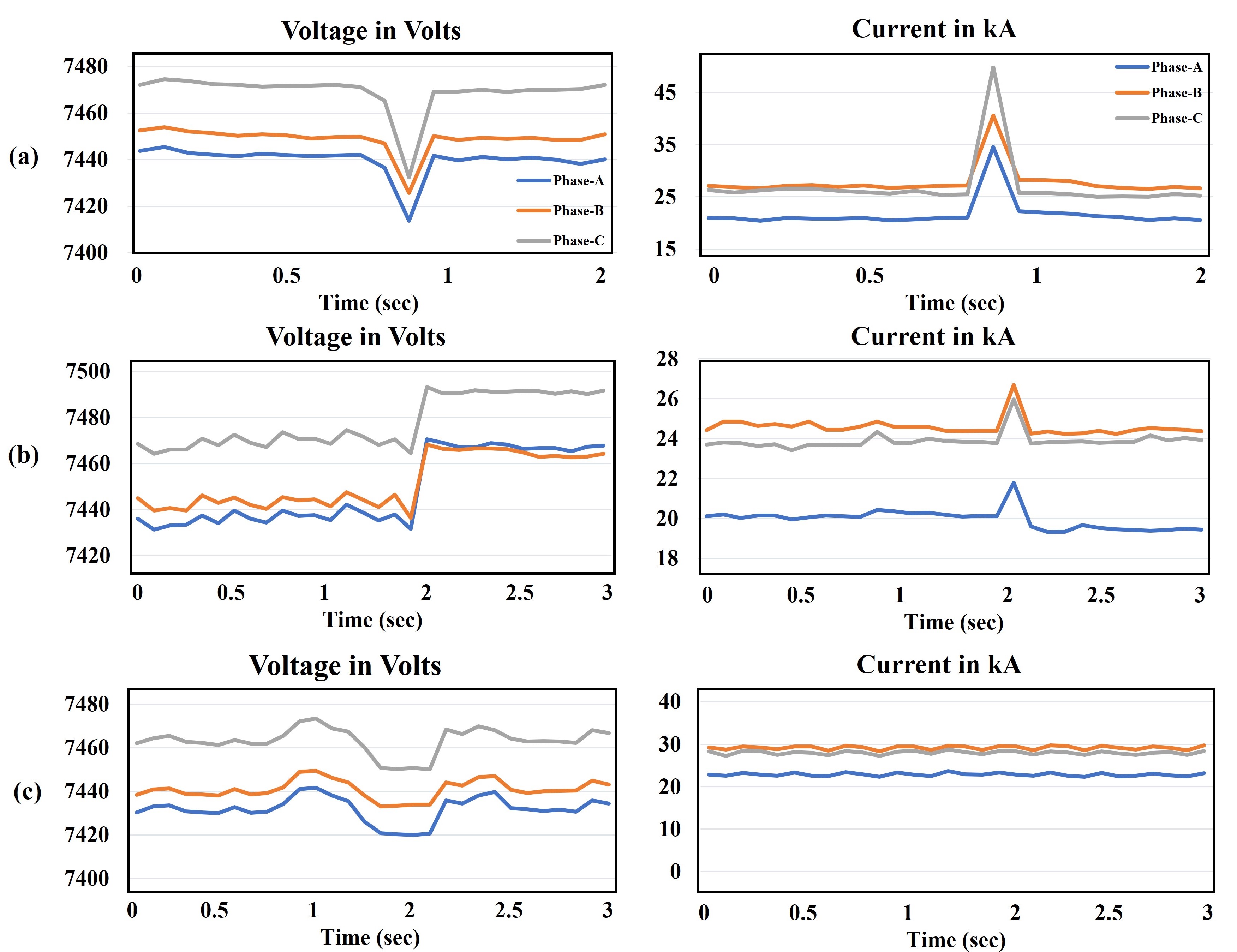}\\
  \caption{Types of anomalies detected in voltage and current measurements; (a) zoom out view of the sudden voltage drop and current increment due to detected inrush current occurred on Day-0, time 18:32:54, (b) zoom out view of capacitor bank switching condition detection with an increase in voltages and sudden increase in currents for all three-phases observed on Day-5, time 10:18:27, and (c) zoom out view of change in the tapping of the transformer event detection with the drop in voltages, no changes in current for all three-phases as observed on Day-3, time 13:22:20.}
  \label{fig:signal}
\end{figure*}

\section{Results and Discussion}
\label{section:Result}

The proposed DynamoPMU algorithm for anomaly detection and prediction is applied to 1 million measurements over 12 days of real-world $\mu$PMU data. For analysis, we considered the current, voltage and power factor measurements of phase-A. Anomaly is detected in the available dataset and metrics are obtained. Then, ten days of data are used for training the prediction model and two days of data are used to test its performance.

\subsection{Anomaly Detection}

All three features' measurements are fed as input to the proposed method. Time-delay coordinates are applied to the time-series measurements with the Koopman operator which provides the first Eigen-time delay coordinates. Then performed singular value decomposition with delay embedding which separates out the linear pattern and forcing operator (rare events). Further, applied the threshold using the three-sigma rule and obtained the anomaly in the data. The obtained result in Fig. \ref{fig:anomaly} shows the anomaly detected in current, voltage and power factor measurements of $\mu$PMU. We can observe that the fluctuating values of the Forcing Operator are obtained, which are further passed through the three-sigma rule for capturing the anomalous events in the dataset. The existence of anomaly is shown with magnitude 1, otherwise, its value is 0. 

\begin{figure}
  \centering
  \includegraphics[width=3.2in]{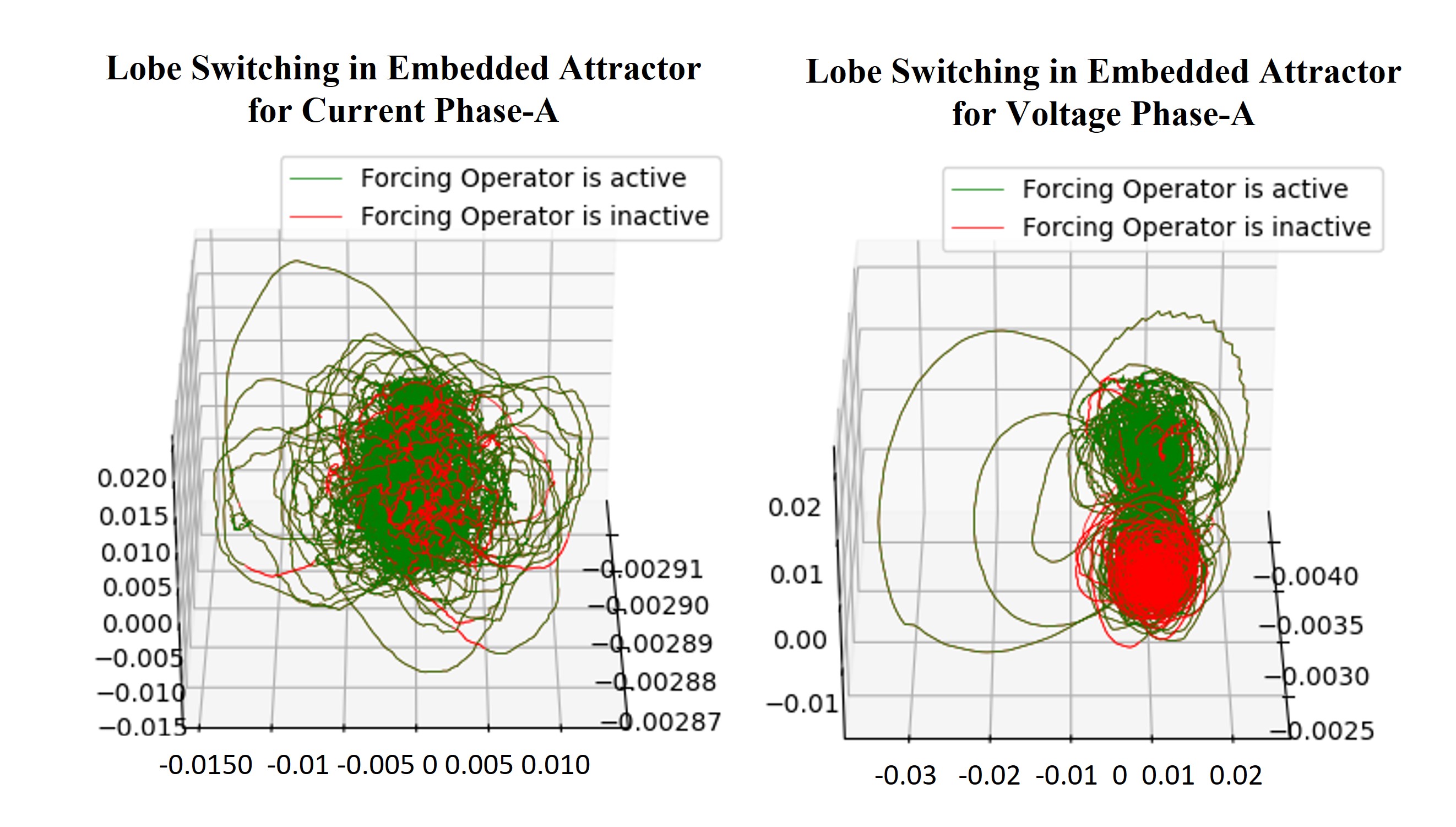}\\
  \caption{Lobe switching in embedded attractor for current and voltage measurements.}
  \label{fig:lobe}
\end{figure}

The proposed method has detected the various events as shown in Fig. \ref{fig:signal} which include capacitor bank switching, inrush current conditions, change in tappings of the transformer, and many other events a few milliseconds before their occurrence. Using the proposed method on Day-0 at time 18:32:54, inrush current condition is detected; on Day-3 at time 13:22:20, change in transformer tappings is detected; and on Day-5 at time 10:18:27, switching capavitor bank condition is detected. Thus, we can use the proposed method for real-time streaming data and detect the events prior to their occurrence such that an alarm or awareness can be raised to deal with them. Also, the lobe switching in the embedded attractor for current and voltage is shown in Fig. \ref{fig:lobe}, which represents the active and inactive status of the forcing operator in a 3D view.

\begin{table}
\centering
\caption{Comparison of scores for existing and proposed event detection algorithms}
\begin{tabular}{>{\centering}m{6em} c c c}
    \toprule
    \multicolumn{1}{>{\centering}m{6em}}{Metrics}  &
    \multicolumn{1}{>{\centering}m{6em}}{Statistical Method\cite{event1}} & \multicolumn{1}{>{\centering}m{6em}}{GAN\cite{event1}} & \multicolumn{1}{>{\centering}m{6em}}{DynamoPMU}\\
\midrule

Precision & 0.596 & 0.947 & 0.970 \\
Recall  & 0.259 & 0.943 & 0.984  \\
MCC  & 0.386 & 0.955 & 0.977  \\
Accuracy & NA & NA & 0.999 \\ 
AUC & NA & NA & 0.992 \\
F1 & NA & NA & 0.977 \\
\bottomrule
\end{tabular}
\label{tab:anomaly_score}
\end{table}

The performance of the proposed method is compared with the existing method using metrics which help in analyzing the performance of the anomaly detection algorithm. From Table \ref{tab:anomaly_score} we can observe that the proposed method outperforms in comparison to the statistical method as well as GAN. As mentioned in \cite{event1}, the MCC score is an important score to check the performance of anomaly detection algorithms because it acquires all four categories of classification into consideration. Here, we can observe that the MCC score is improved significantly in comparison to the GAN. We obtained an MCC score value of 0.977 with the proposed method which shows that it has outperformed in detecting the anomalies. The obtained accuracy for the proposed model is 0.999 which is very high and suggest that the proposed model has effectively detected the anomalies in the dataset. Also, the other score's values are significant in comparison to other methods which shows the superiority of the proposed algorithm. The proposed methodology can be implemented on real-time streaming data as it does not require any prior training data.

\begin{figure}
  \centering
  \includegraphics[width=3.4in]{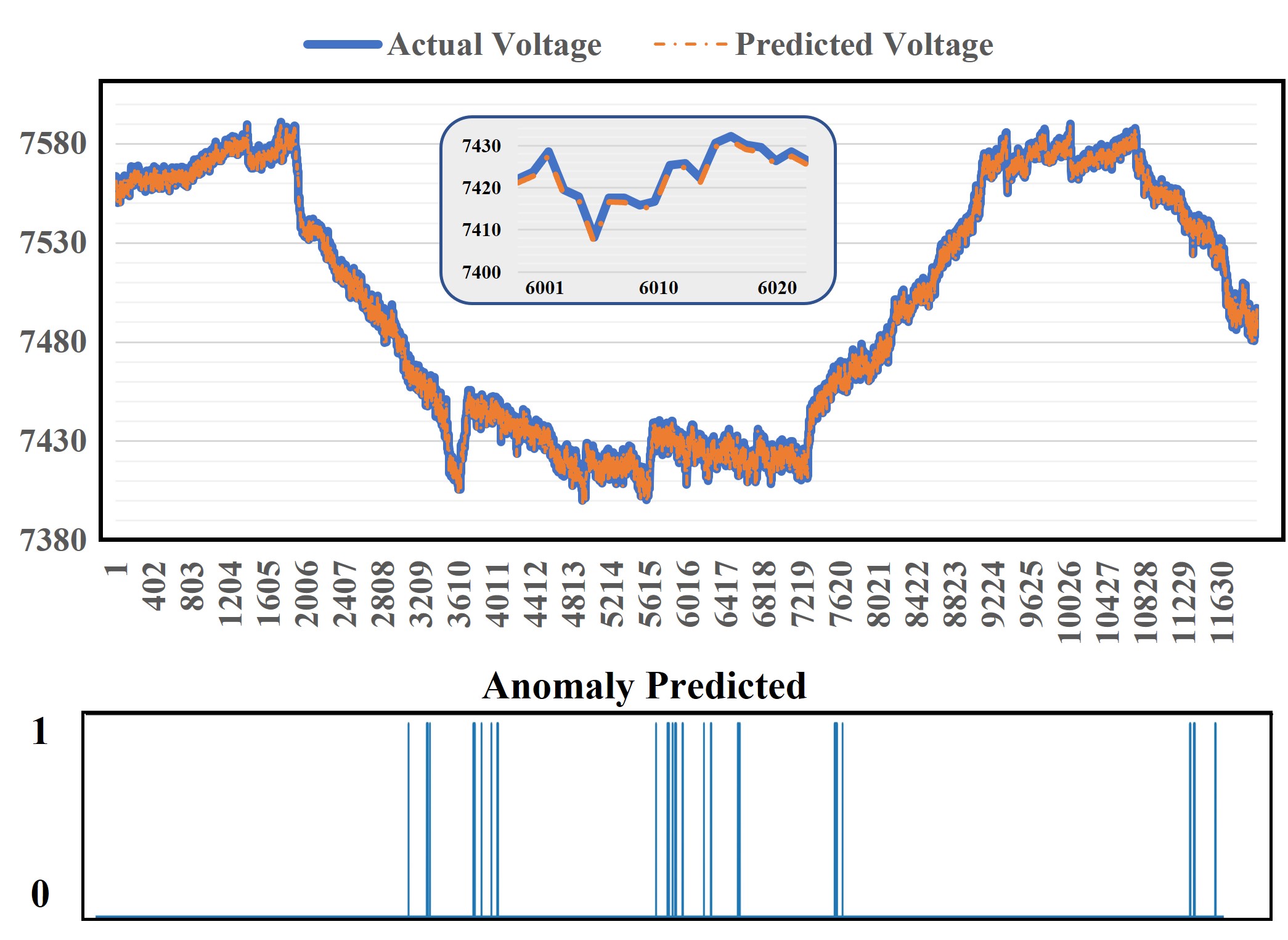}\\
  \caption{Prediction of the voltage measurements with anomaly detection using DynamoPMU for two days and zoom out the voltage signal from 6001-6020 to visualize the variation in actual and predicted voltage.}
  \label{fig:voltage_pred_anomaly}
\end{figure}

\subsection{Anomaly Prediction}

The governing equations for the system, are utilized for the prediction using sparse regression i.e., SINDY model. For training the prediction model we used 10 days and 13 hours of measurement data and predicted the measurements for the next two days. Further implemented the DynamoPMU to the predicted data and obtained the predicted anomaly which would facilitate the situational awareness of the electrical distribution grid. From Fig. \ref{fig:voltage_pred_anomaly}, we can see the predicted voltage for two days using the proposed algorithm and we can say that actual and predicted voltage measurement follows the same pattern. To visualize it more effectively, we have zoomed out and shown the data measurements from 6001-6020 data points in Fig. \ref{fig:voltage_pred_anomaly}. Further, we evaluated the performance of the prediction model using metrics, from Table \ref{tab:prediction} we can observe that the obtained values shows the effectiveness of the proposed algorithm. The $R^2$ and expected variance scores values are near 1 which shows that the prediction of the measurement data is done effectively and we can rely on the model.

\begin{table}
\centering
\caption{Prediction Scores using DynamoPMU}
\begin{tabular}{>{\centering}m{6em} c c c}
    \toprule
    \multicolumn{1}{>{\centering}m{6em}}{Metrics}  &
    \multicolumn{1}{>{\centering}m{6em}}{Scores Values} \\
\midrule

R2 Score & 0.9997  \\
RMSE  & 0.0100   \\
Variance  & 0.9999   \\
MAE & 0.0100  \\ 

\bottomrule
\end{tabular}
\label{tab:prediction}
\end{table}

For real-time implementation for the detection of anomalies, the proposed method is suitable to be integrated into the microcontroller of a $\mu$PMU device or an AI Accelerator attached to it that brings situational awareness and gives an alarm signal to avoid unnecessary trippings. Also, the algorithm is flexible to be integrated into the edge or cloud computing devices for anomaly prediction in the long term.

\section{Conclusion}
\label{section:Conclusion}

A novel unsupervised physics-informed dynamical approach is proposed called DynamoPMU, to detect the anomaly in real-time and simultaneously predict the events in power distribution systems. The proposed framework, based on Koopman theory and Hankel matrix, reconstructs a complex system in the linear intrinsic space where the nonlinear dynamics are expressed in a linear model and intermittent forcing operator. This intermittency represented anomalous events, which were scrutinized and detected in real-time without prior knowledge. We also obtained the governing equations which were utilized to predict future occurring events that help us in handling the events to avoid any damage to the assets and human lives. Sparse identification of the nonlinear dynamical regression model was used for simultaneous event detection. The DynamoPMU is implemented on the online available $\mu$PMU dataset and in terms of performance matrices it was observed that we achieve better scores in comparison to the existing algorithms for anomaly detection as well as prediction of the events. 

In future work, anomaly detection could be utilized to monitor the asset's health diagnostics and distribution-level oscillation analysis. The proposed method would serve as a motivation for research to implement physics inspired dynamical approach on electrical distribution grids for other real-time applications.

\bibliographystyle{IEEEtran}
\bibliography{main}

\end{document}